# Observations Of Extrasolar Planet Transit At The Bosscha Observatory

*R. Satyaningsih[1], B. Dermawan[1], T. Hidayat[1], S. Siregar[1], I. Radiman[1], A. Yamani[1]*
[1]*Bosscha Observatory and Department of Astronomy, FMIPA, ITB*

*Abstract*

*Since its first discovery, most extrasolar planets were detected using radial velocity (RV) method. However, the RV method does not provide all parameters required to characterize a planetary system. Recently, Charbonneau et al. (2000) and Brown et al. (2001) have shown that the RV planet orbiting HD 209458 can be observed using transit method yielding some additional information. As pointed out by Castellano (2004), this method can be undertaken using small aperture telescopes and inexpensive CCDs. We report here new observations of planetary transit in HD 102195 and HD 209458 performed at the Bosscha Observatory since March 2006. Some preliminary results will be presented.*

## I. Introduction

Since 1995, more than 200 extrasolar planets orbiting solar-like stars have been discovered. Most of these planets were detected using the radial velocity (RV) method. However, this method does not provide us information about the inclination of the planetary orbit to the line of sight from Earth. Without this information we cannot deduce the true mass of the planet. Using the transit method we could measure this parameter. Each planet found by the RV method has a probability of transiting its parent star, mainly depending on the orbital period of the planet. The shorter the planet's period, the higher the transit probability is.

During the transit, the brightness of the parent stars will reduce. Although this effect is very faint, it is feasible to measure the relative changes in brightness using inexpensive (ordinary commercial-grade) CCD attached to a small aperture telescope (Castellano 2004).

Furthermore, transitsearch.org, a cooperative observational effort, allows experienced amateur astronomers and small college observatories to discover transiting extrasolar planets. It provides dates and times when transits are expected to occur.

In this paper, we present observations of the photometric dimming of HD 102195 and HD 209458 performed at the Bosscha Observatory since March 2006 based on dates and times provided by transitsearch.org. It is important to point out that no photometric transit search has yet been reported for HD 102195 system, whereas HD 209458 system was the first system known has transiting extrasolar planet (Charboneau et al. 2000).

## II. Observation

Observations of HD 102195 have been made for four nights (March 31, April 4, May 3 and 7, 2006). However, due to bad weather conditions, reliable data were obtained only on May 3, 2006. In this observing run, we used 8-inch aperture GAO-ITB RTS telescope and CCD ST-8XME, resulting a ~23′×15′ field of view. As comparison star, we observed HD 102163 which was in the same field of view with the target star. All measurements were made in clear condition (without filter). During the observation, the cirrus clouds and hazy view dominated. After all, this did not cause a serious problem since we concern mainly on differential photometry. This method measures the target star's brightness relative to the comparison one which is in the same field of view.

In order to detect the planet occulting HD 209458, similar observations were performed on August 17, 2006, using the CCD ST-7XME and the 18-inch aperture GOTO telescope. With this combination we only have field of view as wide as 8.8'×5.9', much narrower than the first combination. Therefore, instead of applying the differential photometry, we took standard star frames needed to extrapolate target star's magnitude in absolute photometry system.

Subsequently, observations on August 24 and 31, 2002, were performed using the same GOTO telescope, but with different CCD. i.e., ST-8XME, so we had a wider field of view, that is, 4.4'×3'. In addition, we also collected standard star frames. All measurements in observation of HD 209458 transit were made through a red (Johnson R) filter. In the beginning of 2006 August 24 runs, clouds were present affecting the frames grabbed in the first one hour were not reliable, so we did not analyze them further. Moreover, frames obtained in the first half of the runs were, unfortunately, out of focus. Note that the observation of August 31, 2006, was ended by the clouds so that we did not achieve a full transit session.

## III. Data Reduction and Photometry Analysis

We used IRAF (Image Reduction and Analysis Facility), a standard program in astronomical image processing, to process the frames obtained from each night of observations. The general procedure of our reduction processes over the frames obtained from each run is as follows:

1. We made a master bias obtained by averaging all available bias frames into one frame.
2. Each image frame and dark frame is reduced by subtracting them from the master bias.
3. Master dark is obtained by averaging the dark frames taken with the same exposure time and close to image-grabbing session.
4. We average all flat frames into one frame then divide it by the average value to obtain a master flat.
5. We produced calibrated images by subtracting a master dark obtained with the same exposure time and dividing by a master flat.

These calibrated images were ready to be analyzed using DAOPHOT package in IRAF. We measured instrumental magnitude for each star from the stellar and sky fluxes presented by IRAF's DAOPHOT. It is important to mention that we obtained a low signal-to-noise ratio (S/N), i.e., much lower than 100, for May 3, 2006, observations, that implies low precision. However, we improved it by stacking the calibrated frames taken in the same run/multiexposure (five or ten frames each) into one frame. Accordingly, we obtain a better S/N and we can undertake photometry analysis with a better precision.

## IV. Results and Discussion

To ensure that the target star and the comparison star in each night of observation have similar behavior over the time of observation, we plotted their instrumental magnitude versus the time of observation. We found that they have similar behavior due to the absence of peculiar data point among them. Figure 1 shows this behavior for May 3 observations. It is important to identify this feature to confirm that the dimming of the target star is due to the presence of the planet, not by other reasons. Subsequently, we obtain the lightcurve by plotting the relative brightness as a function of time.

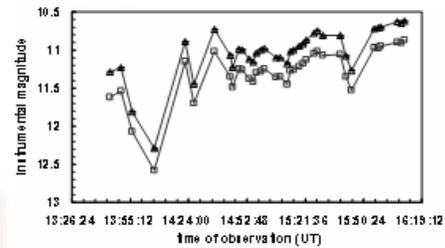

**Fig.1.** The behavior of both the target star, HD 102195 (triangle) and the comparison one, HD 102163 (square) plotted over the time of observation (in UT). It is obvious that they have similar behavior.

For May 3, 2006, observation, the lightcurve is illustrated in Fig.2. Bad seeing probably became the reason for rather large error bar present here. Another error source is poor tracking of the telescope resulting elongated point sources in the frames. Such point source will influence photometry analysis. There will be some area of the point source that is uncovered by the circular aperture so that the flux presented by IRAF does not actually represent the entire flux.

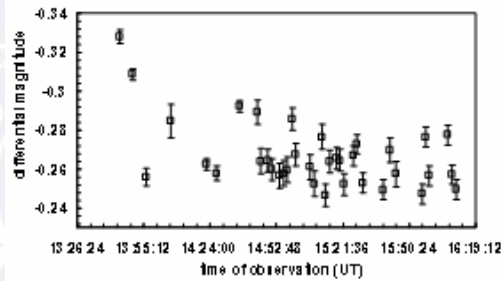

**Fig.2.** Differential lightcurve of HD 102195 observed on May 3, 2006. The largest error is caused by decreasing of the flux due to the clouds.

From these data, it is hard to infer that the HD 102195 has transiting planet although there is such an ingress (brightness decreasing) pattern at the beginning of observation until around 14:24 UT. The data point around 13:55 UT is in doubt to conclude that there is an ingress pattern.

Figure 3 shows the quality of the photometric data we obtained during the observation of HD 209458 transit on the night of 2006 August 17, 24, and 31.

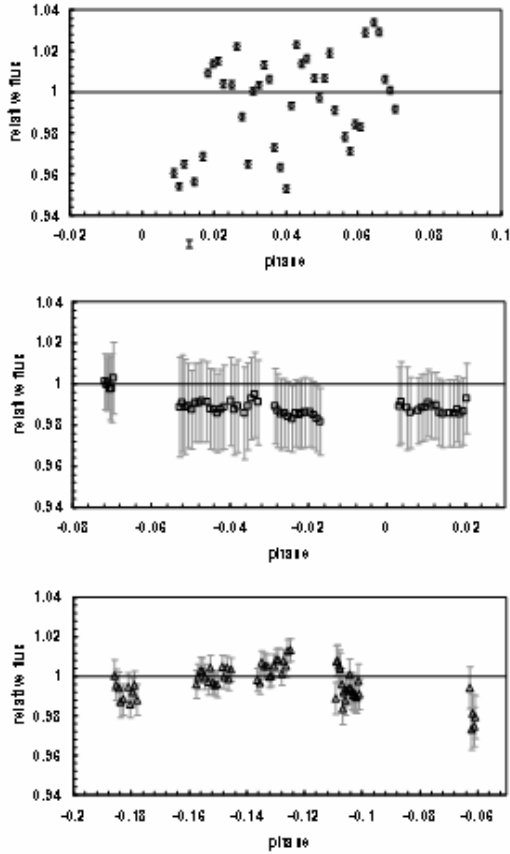

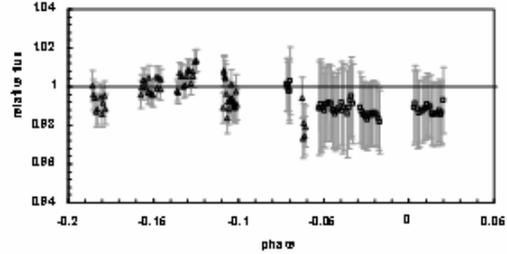

**Fig.4.** The observed transit phenomenon produces a dimming of HD 209458 of less than 2%.

**Fig. 3.** Photometric data obtained during the observation of HD 209458 transit on the night of 2006 August 17, 24, and 31 (upper, middle, and lower panel, respectively).

Note here that we binned three data points gathered on the night of 2006 August 17 into one data point for smoothing. We did the same procedure for the data obtained on the 2006 August 24 Observation, except for about the first 30 minutes data (the "clump" seen on the left of the middle panel in Fig.3). We chose this binning which is reliable enough to represent all data points. Relative flux is gained by comparing the relative fluxes to averaged fluxes over a continuum pattern. Furthermore, we phased our data to a predicted central transit available in transitsearch.org.

To have a distinct visual pre, mid, and post transit curve, we combine two lightcurves in Fig. 3, for 2006 August 24 and 31 observations (see Fig.4). We excluded the data obtained on August 31 because of its scattered data points (see Fig.3) that obscured our transit lightcurve. We can see that during the transit, the amount of light coming from HD 209458 is reduced by less than 2%. It is consistent with the result obtained by Charbonneau et al. (2000).

## V. Concluding Remarks

A transit phenomenon leads to a dimming of the star's light. This dimming can be detected by an inexpensive CCD attached to a small aperture telescope. Our observations still do not confirm whether there is a transit of the planet orbiting HD 102195 or not. More observations obviously are necessary. In addition, HD 102195 is a potential target for southern hemisphere observers since its position in the sky and no photometric results have been reported. In HD 209458 transit observation, our results are consistent with the previous work.

For now we do not provide planetary parameters that can be derived from the transit lightcurve. An appropriate model of fitting to a lightcurve is needed and it will be devoted to our future work.

*Acknowledgments*. This research was supported by "Hibah Penelitian Tim Pascasarjana - Dikti". The authors thank to Bosscha Observatory staffs for their help during the observations.

## VI. References

1) David Charbonneau, Timothy M. Brown, David W. Latham, and Michel Mayor, The Astrophysical Journal 529, 45 (2000).
2) Henden, A. A., Kaitchuck, R. H., "Astronomical Photometry", Van Nostrand Reinhold Company Inc., New York, 1982.
3) M. Irfan, H. Setyanto, I. Ibrahim, Hakim L. Malasan, "Reduksi dan Analisis Data CCD ST-6B di Observatorium Bosscha-Departemen Astronomi ITB Lembang-Jawa Barat", 2003.
4) Timothy M. Brown, David Charbonneau, Ronald L. Gilliland, Robert W. Noyes, and Adam Burrows, The Astrophysical Journal 552, 699 (2001)

5) T. P. Castellano, G. Laughlin, R. S. Tery, M. Kaufman, S. Hubbert, G. M. Schelbert, D. Bohler, and R. Rhodes, JAAVSO 33, 1 (2004).
6) T. P. Castellano and G. Laughlin, "The Discovery of Extrasolar Planets by Backyard Astronomers" (http://www.skywokker.com/transitsearch observingprocedures.htm), 2005

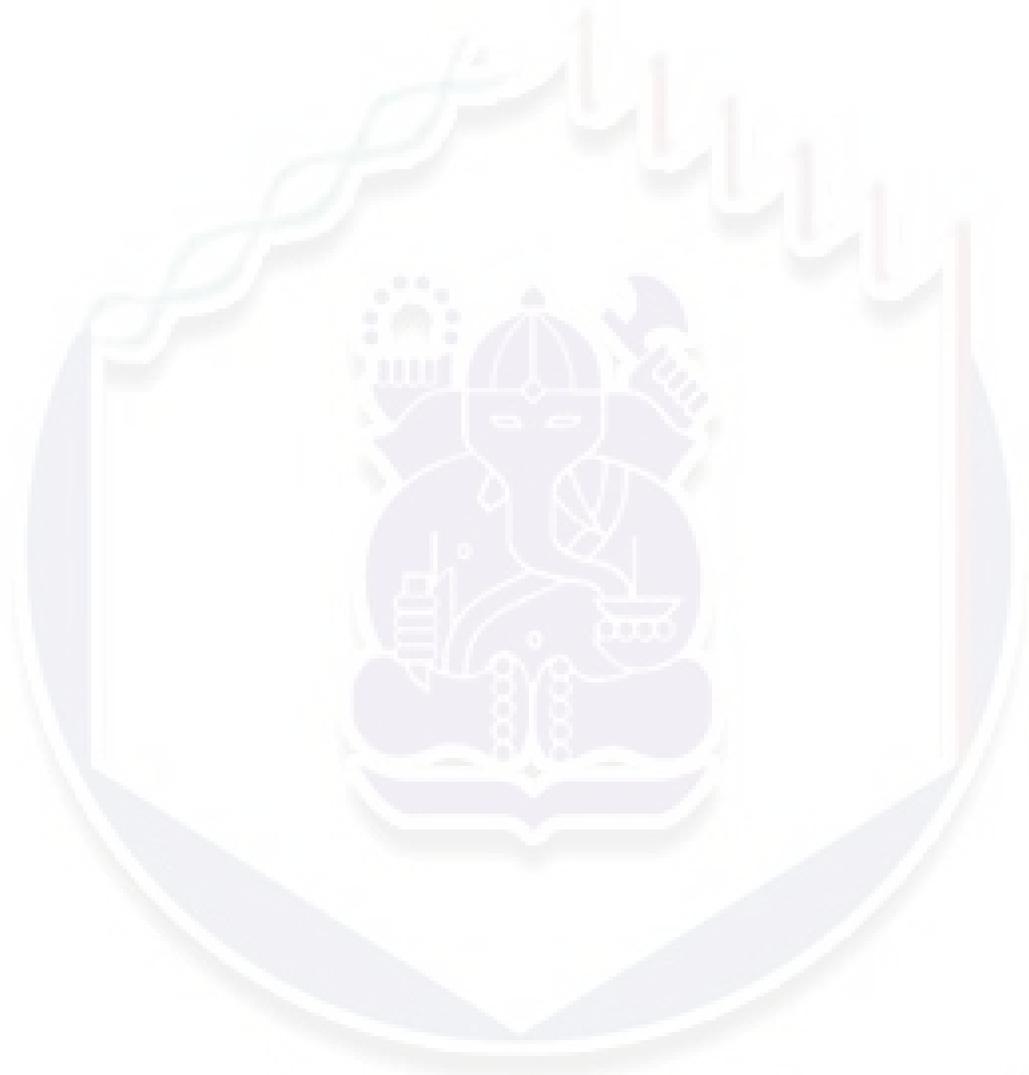